**A Multinomial Model for Comorbidity in England of Longstanding CVD, Diabetes, and Obesity**

Dr Karyn Morrissey

Department of Geography & Planning, University of Liverpool.

Dr Ferran Espuny

Department of Geography & Planning, University of Liverpool.

Dr Paul Williamson

Department of Geography & Planning, University of Liverpool.

## Abstract

From a public health perspective, previous research on comorbidity tends to have focused on identifying the most prevalent groupings of illnesses that demonstrate comorbidity, particularly among the elderly population, already in receipt of care. In contrast, little attention has been paid to possible socio-economic factors associated with increased rates of comorbidity or to the possibility of wider unrevealed need. Given the known relationship between CVD, diabetes and obesity and the strong socio-economic gradients in risk factors for each of the three diseases as single morbidities, this paper uses the Health Survey for England to examine the demographic and socio-economic determinants of each of the seven disease combinations in the English population. Using a multinomial logistic model, this research finds that gender is a significant predictor for all seven disease combinations. However, gender was not as influential as individual age or socio-economic profile. With regard to ethnicity, the black population presents a high obesity, diabetes and diabetes-related comorbidity risk, whilst the Asian population presents a high risk for diabetes and diabetes-related comorbidity but a low risk for obesity and comorbidity. Across the seven disease combinations, risk was lowest for those individuals with a high income (4 out of 7), in-work (4 out of 7), home owning (3 out of 7) and degree educated (3 out of 7). Finally, smokers have a lower risk rate of obesity (and related) than ex-smokers relative to individuals that never smoked (in all cases controlling for all other factors). The important influence of socioeconomic factors has implications for the spatial demand for services and the policy solutions adopted to tackle the increasing prevalence of comorbidity.

### 1. Introduction

Of the estimated 57 million global deaths in 2008, 36 million (63%) were due to non-communicable diseases (NCDs). Population growth and increased longevity are leading to a rapid increase in the total number of middle-aged and older adults, with a corresponding increase in the number of deaths caused by NCDs. In developed countries cardiovascular disease (CVD), diabetes and obesity are major contributors to the overall burden of chronic disease. In 2012, 48% of NCD deaths were caused by CVD and diabetes was directly responsible for 3.5% of NCD deaths (World Health Organisation, 2012). With regard to obesity, the WHO (2011) estimates that approximately 1.5 billion of the global adult population are overweight.

Along with an increased prevalence of CVD, diabetes and obesity as single morbidities, there is a growing body of evidence that individuals are increasingly experiencing two or more of these conditions (Sachdev et al., 2004; Valderas et al., 2009). Comorbidity refers to one or more chronic diseases among people with an index-disease (the primary disease of interest) (Gijsen et al., 2001; Valderas et al., 2009; Congdon, 2010). Responding to increased rates of comorbidity and the knowledge that comorbidity significantly increases mortality rates and decreases functional status and quality of life, the study of comorbidity has increased over the last decade (Islam et al., 2014). With regard to CVD, two of the major risk factors are obesity and diabetes (Nicholls, 2013; Congdon, 2010), and the close link between obesity and diabetes is well established (Balluz et al., 2008). Increases in the prevalence of obesity are a major factor in the growth of diabetes (Congdon, 2010), whilst CVD is listed as the cause of death for 65% of persons with diabetes (Congdon, 2010).

From a public health perspective, previous research into comorbidity tends to have focused on identifying the most prevalent groupings of illnesses that demonstrate comorbidity (Sachdev et al., 2004), particularly among the elderly population (see for example research by Carey et al., 2013; Salive, 2013 Fontin et al., 2007; Marengoni et al., 2011; Gijsen et al., 2001). Indeed, in compiling a literature review on previous research on comorbidity it is difficult to find research that is not limited to the elderly and conceptualises the need for research on comorbidity within the context of the global aging population. However, such a focus ignores the accepted empirical evidence that an individual's health is the outcome of multifaceted processes broader than age, genetics or initial health status alone (Morrissey et al., 2013). Health clinicians, policymakers and researchers now appreciate the role individual level factors have on physical health regardless of age. As such, the multifactorial model of disease causation (Williams 2003; Shim 2002), whereby socio-economic circumstances in conjunction with demographic and genetic factors are seen as the key determinants of individual health outcomes, has become the dominant conceptual framework underlying the analysis of ill health in the social sciences (Morrissey et al., 2013; Williams 2003; Shim 2002).

Socioeconomic status influences health outcomes through a number of mechanisms operating across an individual's life course (Kavanagh et al., 2010). As outlined by Kavanagh et al., (2010), childhood socio-economic circumstances may influence childhood nutrition, behaviours and illness. Whilst in adult life, socio-economic circumstances could influence biological risk factors through a range of factors including health behaviours, psychosocial conditions and access and uptake of health services. Indeed, the WHO (2005) estimates that at least 80% of premature CVD and diabetes could be prevented through modifiable individual risk factors whose prevalence are associated with socio-economic status (Kavanagh et al., 2010) such as healthy diet, regular physical activity and no tobacco use. The existence of this inverse relationship between socioeconomic status and the incidence or mortality rates has been demonstrated for many health outcomes (Zhang et al., 2011; Charlton et al, 2013), including and of particular interest to this paper; CVD (Kavanagh et al., 2010), diabetes (Kavanagh et al., 2010) and obesity (Procter et al., 2008) as single morbidities. Thus, similar to single morbidities, this paper argues that co-morbidity is the result of complex interrelationships between differing health conditions and shared demographic, socioeconomic and environmental risk factors.

From a methodological perspective, the data used in this paper are based on a nationally representative population sample, the Health Survey for England. Research to date on co-morbidity has focused on clinical records; GP, community care and hospital data (Charlton et al, 2013; Valderas et al., 2009; Mulle & Vaccarino, 2013). However, clinical record based data is a source of revealed preference data; only individuals who have actively sought medical attention are included in these datasets. This overlooks those individuals who have co-morbidities but have not sought medical attention. Research on health service access has found a variety of reasons, closely linked to an individual's socio-economic circumstances, such as cost, opportunity cost of time spent away from work or the home, distance and lack of transport that cause individuals not to access health services even when they are ill (Joseph and Phillips, 1984; McLafferty, 2003; Neng et al., 2012). Arguably it may be the characteristics and life circumstances of individuals that are ill and have not sought medical attention that are of most interest to public health research (Neng et al., 2012; Kavanagh et al., 2010).

Previous research on revealed preferences has demonstrated a clear socioeconomic gradient between CVD, diabetes and obesity outcomes as single morbidities (Kavanagh et al., 2010; Bajekal et al., 2013), the co-morbidity of CVD and diabetes (Agardh et al. 2004; Kahn et al. 2008) and diabetes and obesity (Congdon, 2010) and multi-morbidities using administrative data (Charlton et al, 2013). Within this context, this paper argues that there is a clear rationale for investigating the role of individual level demographic and socioeconomic characteristics of comorbidity for the entire English population. The key objective of this paper is to define the population that is at risk of CVD, diabetes and obesity controlling for both demographic and socio-economic characteristics, rather than age profile alone. Using the Health Survey for England; a population based dataset; this paper examines a wider set of demographic and socio-economic factors that may be associated with the comorbidity of CVD, diabetes and obesity at the individual level for England.

## 2. Data

The Health Survey for England (HSE) has been carried out annually since 1995 and provides health, demographic, socio-economic and lifestyle information at the individual level. The survey is designed to represent the population living in private households, and thus it excludes those living in institutions. To date, the majority of studies that look at the health status of the population tend to use surveys that are based on private households only (Acik-Topiak, 2012). Acik-Topiak (2012) highlights that using such data underestimates the extent of poor health in the total population as the unhealthiest, which are overrepresented in institutions e.g. residential care homes, are excluded. However, this paper argues that those residing in residential homes or care facilities with comorbidity have their health needs met and clinical data on their comorbidities exist. In contrast, this paper is specifically interested in the general population that may have unmet health service needs. The number of respondents for the HSE was approximately 16,000 up to 2008; however the sample size has been reduced to 8,000 since 2009. The survey contains interviewer weights, which adjust for selection, non-response, and population age/sex and strategic health authority (SHA) profile, the latter using estimates from the Office of National Statistics (ONS). Unless indicated otherwise, the analysis performed in this paper uses the HSE interviewer weights.

In order to study the comorbidity of cardiovascular disease, diabetes and obesity in adults (age 16 years and above) data from 2008 to 2011 were pooled. People who reported having at least one longstanding illness (LSI) were asked to choose the disease(s) they had been diagnosed with according to the classifications in the International Classification of Diseases (ICD 10). The LSI classification provides data on diabetes and CVD for every year, as detailed below, whereas questions on doctor-diagnosed CVD are only collected on particular but non-consecutive years but from all persons, including those not self reporting a LSI. The need for a large sample of individuals with comorbidity across consecutive years to achieve statistical power, meant that doctor diagnosed CVD was not appropriate for this analysis. A comparison of rates of CVD between the LSI pooled data and the doctor diagnosed CVD for 2011 was carried out. The analysis found that 12.07% of the LSI pooled population reported having CVD compared to 13.6% of the population reporting doctor diagnosed CVD. The similarity in prevalence rates between both populations further justified the decision to concentrate on the LSI population.

Among the (weighted) pooled total of 36,894 HSE participants having given a valid response to the longstanding illness (LSI) question, 15,320 reported having a LSI. This represented 41.53% of the HSE pooled sample. The final sample used within this paper consisted of the 10,161 individuals that reported having a LSI and provided a valid response to all the variables of interest (obesity and equivalised income, defined below, being the most problematic variables). As this paper omits up to two-thirds of the HSE data due to item non response, it is necessary to ensure that the observations for the 10,161 remaining individuals are unbiased. Table 1 presents the age-sex distribution of the two samples is compared using the non-parametric two-sample Kolmogorov-Smirnov test (Conover, 1971). It was found that drawing two variables from the same continuous distribution gives a p-value of 0.98, meaning that the null hypothesis can be accepted. Therefore, the subsample used to model comorbidity within this paper preserves the age-sex distribution of the original LSI population.

**Table 1.** The age-sex distributions of the LSI populations before removing non-response (HSE Ratios), and after doing so (Sample Ratios).

| | **Male** | | | |
|---|---|---|---|---|
| **Age** | **16-54** | **55-64** | **65-74** | **75+** |
| **Sample Ratios (%)** | 24.57 | 10.22 | 7.81 | 5.7 |
| **HSE Ratios (%)** | 22.85 | 9.78 | 7.8 | 6.56 |
| | **Female** | | | |
| **Age** | **16-54** | **55-64** | **65-74** | **75+** |
| **Sample Ratios (%)** | 26.86 | 10.07 | 7.89 | 6.89 |
| **HSE Ratios (%)** | 25.13 | 10.03 | 8.41 | 9.44 |

For the purpose of this study, individuals are categorised as having CVD if they reported any of the following diagnosis: 1) Stroke/cerebral haemorrhage/cerebral thrombosis, 2) Heart attack/angina and 3) Other heart problems. The diabetes variable was taken directly from the survey. No differentiation between the two types of diabetes is available in 2011 because of changing patterns in the prescription of insulin therapy at early ages (which was used in previous years to determine the diabetes type). As is typical in studies using the HSE dataset, obesity was approximated using the weight and height measurements provided by the HSE interviewers were to compute a Body Mass Index (BMI) measure. Using the current definition of obesity, individuals with a BMI above or equal to 30 Kg/m$^2$ were defined as obese for the purpose of this paper. The use of BMI as a health indicator or health outcome is a contested issue (Franzosi, 2006; Evan and Colls, 2009). However, other indicators for obesity such as waist-to-hip ratio, waist measure, and hip measures (Franzosi, 2006) were not available in the HSE. Thus, BMI was used as a proxy for obesity. The impact of using a BMI-based measure of obesity, particularly as it relates to different ethnic groups, is further discussed in Section 5.

With regard to the explanatory variables used in this analysis, the individual level demographic and socio-economic variables are presented in Table 2. The demographic variables include age (4 categories), sex, ethnicity and marital status. The socio-economic variables used include equivalised income (by tertiles), employment status (in work or not), third level education (whether an individual has a degree or not), and tenure (using the available levels in the HSE). Finally, smoking was included as a lifestyle variable. Other lifestyle variables included in socio-economic based analysis of CVD, diabetes and obesity include alcohol consumption, diet and exercise. However for this analysis, the HSE did not include data on physical activity and the question on food and diet only relates to consumption on the previous day. Drinking frequency and quantity estimates are also available in the HSE 2011, but initial analysis found that these variables were highly correlated with age. Therefore, no drinking variable was included in the model. Finally, initial analysis using multinomial regression (outlined in Section 3) included the Index of Multiple Deprivation (IMD 2010) as an independent variable to control for area level deprivation. However, the IMD was found to be not significant and therefore was not included in the final model.

**Table 2.** Characteristics of the 2008-2011 longstanding illness English population

| Variables | Level | Weighted Counts | Ratio |
|---|---|---|---|
| **Total** | | 10157.1 | 100.0% |
| **Gender** | Male | **4905.4** | 48.3% |
| | Female | 5251.7 | 51.7% |
| **Age** | 16-54 | **5223.1** | 51.4% |
| | 55-64 | 2061.5 | 20.3% |
| | 65-74 | 1594.1 | 15.7% |
| | 75+ | 1278.4 | 12.6% |
| **Ethnicity** | White | **9484.1** | 93.3% |
| | Black | 163.3 | 1.6% |
| | Asian | 380.0 | 3.7% |
| | Other | 129.7 | 1.3% |
| **Marital Status** | Single | **1562.4** | 15.4% |
| | Married/Civil Partners | 5530.6 | 54.4% |
| | Cohabitees | 1048.2 | 10.3% |
| | Separated/Divorced | 1089.7 | 10.7% |
| | Widowed | 926.2 | 9.1% |
| **Income Tertile** | Lowest | 3549.5 | 34.9% |
| | Medium | 3514.9 | 34.6% |
| | Highest | **3092.7** | 30.4% |
| **In-Work** | Yes | **4836.9** | 47.6% |
| | No | 5320.2 | 52.4% |
| **University** | Yes | **1877.3** | 18.5% |
| | No | 8279.8 | 81.5% |
| **Tenure** | Renting or Free | **3012.5** | 29.7% |
| | Owning | 3618.4 | 35.6% |
| | Buying or Shared | 3526.2 | 34.7% |

| Smoking | Never Regular | **4859.0** | 47.8% |
|---------|---------------|------------|-------|
|         | Currently     | 2217.1     | 21.8% |
|         | Ex-Regular    | 3081.0     | 30.3% |

### 3. Statistical Model: Multinomial Logistic Regression

Using variables that have been found to be significantly associated with each of the three morbidities, this paper uses a multinomial log-linear model to model the demographic and socio-economic determinants of comorbidity (CVD, diabetes and obesity) at the individual level for England. Multinomial log-linear models are used when the dependent variable to be explained is polytomous and categorical, i.e. it has more than two categories with no global order between them (Morgon and Teachman, 1988). Comorbidity is studied using a response variable $y$ with 8 categories, corresponding to the product of all possible combinations of CVD, diabetes, and obesity status for the considered longstanding illness population; concisely: non-obese, non-CVD, non-diabetic ($y = 1$); obese, non-CVD, non-diabetic ($y = 2$); CVD, non-obese, non-diabetic ($y = 3$); diabetic, non-obese, non-CVD ($y = 4$); obese, CVD, non-diabetic ($y = 5$); obese, diabetic, non-CVD ($y = 6$); CVD, diabetic, non-obese ($y = 7$); obese, CVD, diabetic ($y = 8$).

The probability for an individual $i$ to be in the comorbidity category $c$ is modelled under the multinomial log-linear model as

$$P(y_i = c) = \frac{e^{\beta_c X_i}}{1 + \sum_{k=2}^{8} e^{\beta_k X_i}} \qquad (1)$$

where $X_i$ is the vector of values of the independent variables $X$ for individual $i$, and $\beta_c$ is the vector of regression coefficients for the comorbidity category $c$, being the non-morbid category ($y = 1$) selected as reference: $\beta_1 \equiv 0$. All the independent variables are categorised using dummies, so that each $X_i$ is in fact a vector of 0 and 1 values. The easiest way to interpret the fitted multinomial model is to look at its log-odds, which correspond to the logarithm of the relative risks of moving from the reference category ($y = 1$) to a comorbidity category $c$:

$$\log\left(\frac{P(y_i=c)}{P(y_i=1)}\right) = \beta_c X_i \qquad\qquad (2)$$

By denoting by $\beta_c^j$ and $X_i^j$ the components of $\beta_c$ and $X_i$, respectively, the log-odd expression $\beta_c X_i$ expands as a sum of terms $\beta_c^j X_i^j$; the effect of each log-odd coefficient $\beta_c^j$ is additive and must be only considered when $X_i^j$ is non-zero.

Nnet within the R software (Venables & Ripley, 2002) is a statistical package that supports the fitting of multinomial log-linear models. Its function 'multinom' (Venables & Ripley, 2002) was used to fit the described model for different tested sets of explanatory variables and categorisations of those variables. The final model selected was the one with the set of regressors (Table 1) with the best Akaike Information Criterion (AIC), thus giving the best compromise between goodness of fit and complexity (Burnham & Anderson 2004). The final model coefficients and their significance levels are presented in Table 3.

## 4. Results

### *The Prevalence of Comorbidity*

This study focuses on the comorbidity of CVD, diabetes and obesity, using the population of individuals with a longstanding illness (LSI) in the weighted HSE dataset. Table 3 presents the comorbidity profile by age/gender for the English population that report having an LSI. This profile was estimated using temporally pooled 2008-2011 HSE data. The prevalence of single morbidity obesity (obese, no CVD, and no diabetes) was higher for females than for males, at any age; however, male ratios are higher for the majority of the other single morbidities and comorbidities at each age. Prevalence ratios increase with age (age gradient) for the single morbidities CVD and diabetes, and the comorbidities of obesity & CVD, and CVD & diabetes. In contrast, such increase stops at the age of 74 years for the single morbidity obesity, and the comorbidities of obesity & diabetes, and of obesity & CVD & diabetes. For ages above 55, single morbidities represent overall between 35% and 40% of the total LSI population for both genders. Comorbidities as a whole represent between 16% and 20% for men, and a slightly smaller proportion for women, ranging between 12% and 15% of the total LSI population. The considered comorbidity categories are significantly represented and merit the further analysis that follows.

**Table 3.** Prevalence (%) of obesity, cardiovascular disease (CVD), and diabetes by age/gender group among the pooled HSE 2008-2011 longstanding illness English population

|  | Obese | CVD | Diabetes | Obese, CVD | Obese, diabetic | CVD, Diabetic | Obese, CVD, Diabetic |
|---|---|---|---|---|---|---|---|
| **Male 16-54** | 23.1 | 2.9 | 3.6 | 1.7 | 3.7 | 0.1 | 0.5 |
| **Male 55-64** | 25.3 | 8.8 | 5.4 | 7.1 | 6.6 | 0.6 | 1.6 |
| **Male 65-74** | 17.1 | 11.8 | 7.3 | 7.4 | 6.8 | 2.2 | 3.9 |
| **Male 75+** | 11.6 | 19.4 | 8.1 | 9.7 | 3.5 | 4.1 | 1.9 |
| **Female 16-54** | 27.8 | 2.1 | 2.2 | 1.2 | 2.7 | 0.04 | 0.3 |
| **Female 55-64** | 27.7 | 5.2 | 2.5 | 3.8 | 6.5 | 0.4 | 1.2 |
| **Female 65-74** | 27.7 | 7.7 | 4.3 | 4.5 | 5.9 | 1.1 | 2.2 |
| **Female 75+** | 20.1 | 13.8 | 5.5 | 7.3 | 4.0 | 2.3 | 1.4 |

### *Multinomial Logistic Analysis*

Table 4 continues the analysis of the demographic and socio-economic determinants of comorbidity by presenting the results of the multinomial model. The comparison between intercepts in the multinomial model shows that the risk of comorbidity between CVD & diabetes is less probable for the reference group (male, age 16 to 54, white, single, high income, in-work, with a degree, owning their own home and never a regular smoker). From Table 4 one can see that the estimated multinomial model confirms the gender patterns observed for the prevalence ratios in Table 3. Namely, the risk for single obesity is higher for women than for men, whereas the risk for all other single morbidity and comorbidity categories is significantly higher for men than for women. Increasing age gradients are observed with the relative risk of CVD (log-odds 1.16, 1.26, 1.7), diabetes (log-odds 0.63, 0.98, 1.16), the comorbidity of obesity and CVD (log odds 1.34, 1.23, 1.47), the comorbidity of CVD & diabetes (log-odds 2.57, 3.44, 4.16) and the comorbidity of obesity & CVD & diabetes (1.29, 1.6, 0.81). Table 4 observes an increasing significant age gradient for obesity until age 74 years plus. However, interestingly individuals aged 74 years plus have decreased prevalence ratios for obesity (log-odds 0.2, 0.001, -0.41). At the same time, it is interesting to note that controlling for demographic and socio-economic characteristics, Table 4 indicates that the 74 years old plus age group has the lowest levels of obesity. This result may be due to one or more of a combination of three circumstances. Firstly, that the oldest individuals within the group have lower rates of obesity (a cohort effect). Secondly, that obesity lowers life expectancy, through the onset of diseases attributable to being overweight and obesity (Foresight, 2007). Thirdly, as this paper is examining private households only, there may be a effect of transfer to care homes.

Ethnicity was found to be significant for obesity and diabetes as single morbidities; and obesity & diabetes, and CVD & diabetes as comorbidities. Single morbidity diabetes, and the comorbidities of obesity & diabetes, and CVD & diabetes are more likely to happen in black (log-odds 1.78, 1.3, and 1.77, respectively) and Asian (log-odds 1.81, 0.79, 2.26) populations than in the predominant white population. This result is similar to recent research by Tallin et al. (2012), whose study reported diabetes to be the most prevalent in these ethnic groups. Tallin et al. (2012) found insulin resistance and truncal obesity to be important medical determinants for diabetes risk in Indian-Asian and African-Caribbean females. Individuals of Asian ethnicity were found less likely to be obese than the white population (log-odd -0.66). Both populations have a lower obesity risk than the black population (log-odd 0.45). This result, in combination with observed higher diabetes risk, points to the debate on the suitability of the current BMI threshold defining obesity for the Asian population (see e.g. James et al., 2002) but also for the black population. The recent NICE Public Health Guidance (2013) is to use specific BMI thresholds (lower than usual) for both ethnic groups to define obesity only for diabetes prevention, but not for the general purpose of defining obesity. However, to date specific BMI thresholds by ethnicity have not been agreed for practical use.

Several marital status categories were found to have a significant relationship with a number of the morbidities and comorbidities of interest. In all cases, being single was associated with a lower relative risk in comparison to other marital statuses. Concerning obesity, the marital statuses sorted by increasing relative risk are separated or divorced, cohabitees, married/civil partnership, and widowed (log-odds 0.33, 0.39, 0.46, and 0.51, respectively). For CVD, the marital statuses in increasing order of relative risk are married/civil partnership, cohabitees, and widowed (log-odds 0.49, 0.57, 0.67). The obtained marital status effect on the relative risk for the comorbidity of obesity & CVD are consistent with the single morbidity results: being married or in a civil partnership implies a higher risk (log-odd 0.73) than being single and such risk is slightly higher for widowed (log-odd 0.83). Examining the comorbidity of obesity & diabetes, being married/civil partnership, being separated or divorced, and being widowed were found significant (log-odds 0.9, 0.96, 1.12). Being married was also found to be significant (log-odd 0.76) for the comorbidity of obesity & CVD & diabetes.

Examining the socio-economic variables, Table 4 indicates that being in the lowest or medium income tertile (log-odds 0.41, 0.19), not having a degree (log-odd 0.39), and renting (log-odd 0.13) has a significant positive effect on obesity. The National Obesity Observatory (National Obesity Observatory, 2012) has reported similar income and education incidence rates for the English population. Individuals in the lowest and medium income tertile (log-odd 0.39; 0.43), not being in-work (log-odd -0.69), and renting (log-odd 0.31) have a significant positive effect on CVD.

Concerning the comorbidities obesity & CVD, the multinomial model found not being in-work (log-odd 0.58), not having a degree (log-odd 0.49), renting (0.64) and buying (log odds 0.47) were positively significant. For obesity & diabetes, a significant negative relationship was found if an individual was in the lowest or medium income tertile (log-odd 0.51; 0.3) or did not have a degree (log-odd -0.42). Examining CVD & diabetes comorbidity, a positive significant effect was found for not being in-work (log odds 1.56). With regard to the comorbidity of obesity & CVD & diabetes, it was found that the lowest income tertile (log-odd 0.97), not being in-work (log-odd 1.13), being a home renter or buyer (log-odds -0.93, 0.27) had a significant positive effect on comorbidity.

In terms of analysis, an issue with multinomial models is the amount of information provided by one model. Given the specific interest of this paper, Table 5 presents a map of the socio-economic factors and the combinations of morbidities and comorbidities that they have the strongest significant effects on, controlling for the demographic and socio-economic variables included in the model. From Table 5, it becomes clear that the income gradient presented in Table 4 has the greatest effect on obesity based combinations of comorbidity, obese, CVD & diabetes, obese & diabetes and obese & CVD. Employment status (not in work) has the greatest significant effect on CVD based combinations of comorbidity and CVD as a single morbidity. Not having a third level degree was found to have the strongest impact on obesity based comorbidities and obesity as a single morbidity. Finally, tenure (owning your own home), similar to employment status, has the greatest significant effect on CVD based combinations of comorbidity and CVD as a single morbidity. These results are interesting and indicate that the socio-economic factors associated with differ combinations of comorbidities are not homogenous. However, the set of socio-economic factors included in this analysis have a stronger relationship with obesity and CVD as both single and comorbidities, compared to diabetes.

**Table 4.** Multinomial logistic model of comorbidity of obesity, cardiovascular disease (CVD) and diabetes among the pooled 2008-2011 longstanding illness population in England (the log-odd coefficients of the model and their significance levels are shown). The reference group is: Male, Age 16—54, White, Single, High Income, In-work, Degree, Owning, and Never Regular Smoker.

| | Obese | CVD | Diabetic | Obese, CVD | Obese, diabetic | CVD, Diabetic | Obese, CVD, Diabetic |
|---|---|---|---|---|---|---|---|
| **Reference Group** | -1.89 ** | -4.21 ** | -3.56 ** | -5.16 ** | -4.36 ** | -9.6 ** | -7.65 ** |
| **Female** | 0.16 ** | -0.58 ** | -0.59 ** | -0.57 ** | -0.29 ** | -0.86 ** | -0.59 ** |
| **Age 55-64** | 0.2 ** | 1.16 ** | 0.63 ** | 1.34 ** | 0.79 ** | 2.57 ** | 1.29 ** |
| **Age 65-74** | 0 | 1.26 ** | 0.98 ** | 1.23 ** | 0.62 ** | 3.47 ** | 1.6 ** |
| **Age 75+** | -0.41 ** | 1.7 ** | 1.16 ** | 1.47 ** | -0.02 | 4.16 ** | 0.81 * |
| **Ethnicity: Black** | 0.45 * | -1.18 | 1.78 ** | -0.56 | 1.3 ** | 1.77 ** | 0.51 |
| **Ethnicity: Asian** | -0.66 ** | 0.27 | 1.81 ** | -0.26 | 0.79 ** | 2.26 ** | -0.52 |
| **Ethnicity: Other** | 0 | -0.14 | 0.85 * | 0.21 | 0.35 | -2.57 | 0.9 |
| **Married/Civil Partnership** | 0.46 ** | 0.49 ** | 0.31 | 0.73 ** | 0.9 ** | 0.81 | 0.76 * |
| **Cohabitees** | 0.39 ** | 0.57 ** | 0.26 | 0.37 | 0.49 | 1.11 | 0.38 |
| **Separated/Divorced** | 0.33 ** | 0.29 | -0.01 | 0.47 | 0.96 ** | 0.48 | -0.14 |
| **Widowed** | 0.51 ** | 0.67 ** | 0.33 | 0.83 ** | 1.12 ** | 0.62 | 0.63 |
| **Income: Low** | 0.41 ** | 0.39 ** | 0.13 | 0.26 | 0.51 ** | 0.77 | 0.97 ** |
| **Income: Medium** | 0.19 ** | 0.43 ** | 0.13 | 0.1 | 0.3 * | 0.55 | 0.63 |
| **Not In-Work** | 0.03 | 0.69 ** | 0.14 | 0.58 ** | 0.2 | 1.56 * | 1.13 ** |
| **Education: No Degree** | 0.39 ** | 0.1 | 0.1 | 0.49 ** | 0.42 ** | 0.2 | 0.79 |
| **Renting or Free Ownership** | 0.13 | 0.31 ** | -0.04 | 0.64 ** | 0.21 | 0.42 | 0.93 ** |
| **Buying or ½ Rent/Mortgage** | 0.21 ** | 0.17 | -0.12 | 0.47 ** | 0.09 | 0.21 | 0.27 |
| **Smoking Currently** | -0.42 ** | 0.03 | -0.03 | -0.38 * | -0.4 ** | 0.27 | -0.78 * |
| **Smoking Ex-regularly** | 0.16 ** | 0.21 * | 0.35 ** | 0.5 ** | 0.27 * | -0.07 | 0.67 ** |

*Significant at the 0.05 level, **Significant at the 0.01 level.

**Table 5 Map of the Strongest Socio-economic factors for each disease combination**

| Socio-economic Factor | Strongest Significant Effect | Second Strongest Significant Effect | Third Strongest Significant Effect |
|---|---|---|---|
| Income gradient | Obese, CVD & Diabetes | Obese & Diabetes | Obese & CVD |
| Not In Work | Obese, CVD & Diabetes | Obese & CVD | CVD |
| Education: No degree | Obese & CVD | Obese & Diabetes | Obese |
| Tenure | Obese, CVD & Diabetes | Obese & CVD | CVD |

Finally, examining the smoking variable, the multinomial model found a higher disease risk for ex-regular smokers and a lower risk for current smokers in all the obesity-related morbidities and comorbidities (see Table 4 for the log-odd values). Given the known health risks associated with smoking, this was deemed a surprising result. To examine this relationship further, a simple logistic model was run for BMI using the smoking status as independent variable (categories smoking currently, smoking ex-regularly, never regularly). The logistic model showed that average BMI is lower for current smokers than for non-smokers, whilst BMI is highest for ex-smokers. These relationships are significant. An increase in weight when quitting smoking has been reported elsewhere (Aubin et al. 2012; Alley et al., 2010), and this paper hypothesises that the decreased obesity risk for current smokers compared to ex-smokers is associated with such weight gain. Being an ex-regular smoker is associated with a higher risk for single morbidities CVD and diabetes (log-odds 0.2 and 0.35, respectively). Unfortunately, unlike the obesity variable, no conclusion can be extracted for current smokers, since the corresponding model coefficients were not significant. However, it is important to note that numerous medical studies show that quitting smoking reduces the risk for CVD, independently of diabetes (see e.g. Clair et al. 2013).

### 5. Discussion

Numerous studies have analysed the effects of individual level socio-economic status on cardiovascular health, diabetes and diabetes and obesity. As such, the socio-economic determinants of each of these morbidities are well established at the national level and include low socio-economic status and unemployment (Fone et al., 2013; Kavanagh et al., 2010; Congdon, 2010). However limited research to date has examined the association between socio-economic status and the comorbidity of CVD, diabetes and obesity in itself. Previous clinical research on comorbidity tends to have focused on identifying the most prevalent and prognostically important illnesses that tend to demonstrate comorbidity (Sachdev et al., 2004), whilst both clinical and population based research has tended to focus singly on the elderly population. Widening the population to include all individuals aged sixteen years and above, this paper presents the first socio-economic based research on the comorbidity of CVD, diabetes and obesity in the UK. Using a population dataset of private households, the HSE, it was found that controlling for a host of demographic and socio-economic factors the individual level covariates associated with different morbidities and combinations of comorbidities are not homogenous (Table 4). Demographic and socio-economic factors vary in significance and magnitude of association across each morbidity and comorbidity combination.

Examining the demographic variables, this analysis indicates that gender and marital status are significant predictors of all seven disease combinations, except for obesity alone. However, these two explanatory variables are not as strongly significant as other variables. The multinomial logistic model found that increasing age gradients had a significantly higher risk for CVD, diabetes, and their co-morbidity. With regard to ethnicity, the black population presents a high risk for diabetes (and diabetes-related co-morbidity), whilst the Asian population presents a risk for high diabetes risk but low risk for obesity. However, whilst BMI is now the dominant tool used in quantitative analyses of obesity, it is important to note the discussions regarding the limits of the BMI (Nicholls, 2013), including the recommendation to apply ethnic specific BMI thresholds. With regard to obesity and comorbidity, Bell et al., (2002) state that ethnic-specific definitions of obesity has been hindered by a lack of data clarifying whether or not obesity-related comorbid conditions occur at different levels of BMI in different ethnic groups. Thus, more research needs to be conducted in ethnically diverse populations if the relationships between BMI, body fat and chronic disease are to be better defined (Katzmarzyk, 2011; Nicholls, 2013).

With regard to socioeconomic status, having controlled for demographic and lifestyle factors, the risk for each of the seven disease categories was lowest for the individuals with an equivalised income in the highest tertile, in-work, owning their own home and having a degree, of which not in-work had the greatest association. With regard to smoking, smokers have a lower risk rate of obesity (and related comorbidities) than ex-smokers relative to individuals that never smoked. The strong associations between lower socio-economic status (income, education, employment status and tenure) and comorbidity, having controlled for demographic characteristics, reinforces the need to concentrate health promotion and health policy efforts on reducing social inequalities rather than behavioural characteristics alone (Kavanagh et al., 2010).

Following Frolhlich and Polvin (2008), this paper argues that interventions that attempt to alter some of society's behavioural norms, for example, the current example of banning smoking in public places, assumes that everyone's risk exposure is reduced by the same amount, regardless of one's initial position in the risk exposure distribution. In reality, socio-economic position has been repeatedly shown to influence individual's behavioural characteristics (Kavanagh et al., 2010;

Frolich and Potvin 2008). Thus, given the significant association between socio-economic characteristics and comorbidity demonstrated in this paper, this paper argues that a focus on vulnerable populations is complementary to a population approach and necessary for addressing social inequalities in health. As noted by Frolich and Potvin (2008), the notion of vulnerable populations differs from that of populations at risk in that a population at risk is defined by a higher measured exposure to a specific risk factor. In contrast, a vulnerable population is a subpopulation which, because of shared social characteristics, is commonly exposed to contextual conditions that places it at a higher risk than the rest of the population. The findings of this paper indicate that individuals in lower socio-economic groups across England, controlling for demographic factors may be defined as vulnerable groups. Thus, we argue that inequalities in health outcomes are unlikely to change without attention being paid to the generators of socio-economic inequalities. With regard to health policy, these findings indicate that solutions to redressing health inequalities may lie outside the health sector (Kavanagh et al., 2010; Frolich and Potvin 2008); thus, calling for a more inter-disciplinary approach to public health provision.

Whilst the research presented in this paper focused on individual level factors, future research on comorbidity must include both individual and contextual environmental factors. Place of residence is strongly patterned by social position; neighbourhood characteristics may be important contributors to health disparities (Diez Roux and Mair, 2010). There is a growing appreciation of the role that contextual environmental factors or "neighborhood" effects play on physical health (Diez Roux and Mair, 2010). Neighbourhoods have both physical and social attributes that may influence health. Physical attributes include access to goods and services, green space, and availability of alcohol and tobacco outlets; social attributes include community unemployment, segregation, social capital, civic participation, and crime (Diez Roux and Mair, 2010). Thus, the next step in this research is to incorporate spatial referencing within the HSE via a spatial microsimulation algorithm (Morrissey et al., 2013; Clarke et al., 2014) and the Census of Population 2011 to help understand how rates of comorbidity are represented across England. That said, this paper found that the IMD, a spatially referenced social and economic indicator for each lower super output area in England, did not have a significant association with any of the morbidities or comorbidities within the model. However, creating a spatially representative demographic, socio-economic and co-morbidity profile for the population of England would allow research to highlight comorbidity hotspots in relation to current health supply facilities. Such an analysis would therefore highlight areas with potential unmet health service needs (Morrissey et al., 2008).

## 6. Conclusion

Previous research on comorbidity has focused on identifying the most prevalent groupings of illnesses that demonstrate comorbidity (Sachdev et al., 2004), particularly among the elderly population. This paper argues that such a focus ignores the accepted empirical evidence that an individual's health is the outcome of multifaceted processes rather than age or initial health status alone (Morrissey et al., 2013). Instead, the relative risk of comorbid health outcomes should be seen as being shaped from an early age by individual level circumstances. Extending previous research on the comorbidity of CVD, diabetes and obesity, this paper found that just as for to CVD, diabetes and obesity as single morbidities, socio-economic factors are an important determinant of comorbidity in England. This suggests that increased emphasis needs to be placed on determining the socio-economic determinants co-morbidity both in England and internationally. However, as noted by Kavanagh et al., (2010), to change socio-economic outcomes in health outcomes, research concentrating on individual risk factors needs to be conducted in tandem with research on the generators of socio-economic inequalities. Thus, both in the UK and internationally the public health agenda requires numerous agencies working together in tandem to prevent inter-generational socio-economic factors determining an individual's health outcome.

With regard to the provision of health services, this study indicates that in England single disease management approach is no longer suitable for a large number of patients. Since comorbidity is significantly related to increased levels of mortality and decreased functional status and quality of life, health care should shift its focus from specific diseases, to multiple pathologies, worsening functional status, increasing dependence of care and the increased risk of mental and social problems (Gijsen et al., 2001; Islam et al., 2014). Furthermore, given the socio-economic gradient observed in this paper and previous international research (Kavanagh et al., 2010) interventions to reduce comorbidities should be tailored to the unique risk profile and needs of high-risk communities (Rodriguez et al., 2013).

**Acknowledgements**

Funding for this research was provided by the ESRC Secondary Data Analysis Initative, Grant Number ES/K004433/1.